\documentclass{ws-ijmpd}

\begin{document}

\markboth{H. Ito, M. Kino, N. Kawakatu and S. Yamada}
{Non-Thermal Emissions from  Shells Driven by  AGN Jets}

%
\catchline{}{}{}{}{}
%

\title{NON-THERMAL EMISSIONS FROM SHOCKED SHELLS DRIVEN BY POWERFUL AGN JETS
  }

\author{HIROTAKA ITO}

\address{Research Center for the Early Universe, School of Science, the
University of Tokyo, \\
 Bunkyo-ku, Tokyo 113-0033, Japan \\
ito@resceu.s.u-tokyo.ac.jp}

\author{MOTOKI KINO}

\address{National Astronomical Observatory of Japan, \\
 Mitaka, Tokyo 181-8588, Japan}

\author{NOZOMU KAWAKATU}

\address{Department of Physics, University of Tsukuba, \\
  Tennodai, Tsukuba 305-8577 Japan }

\author{SHOICHI YAMADA}

\address{Science and Engineering, Waseda University, \\ 
Shinjuku, Tokyo 169-8555, Japan}

\maketitle

\begin{history}
\received{Day Month Year}
\revised{Day Month Year}
\comby{Managing Editor}
\end{history}

\begin{abstract}
 We explore the emissions by accelerated electrons in
 shocked shells driven by jets in active galactic nuclei (AGNs). 
 Focusing on powerful sources which host luminous quasars, 
  the synchrotron radiation 
 and inverse Compton (IC) scattering of various  photons
 that are mainly produced in the  core are 
 considered as radiation processes.
%
%
 We show that
 the radiative output is dominated by the IC emission for compact 
 sources ($\lesssim 30{\rm kpc}$), whereas the synchrotron radiation 
is more important for larger sources.
 It is predicted
 that,
for
 powerful sources ($L_{\rm j} \sim 10^{47}{\rm ergs~s^{-1}}$),
 ${\rm GeV}-{\rm TeV}$ gamma-rays
 produced via the IC  emissions 
 can be detected by the Fermi satellite and modern Cherenkov 
telescopes such as MAGIC, HESS and VERITAS if the source is compact.
%

\end{abstract}


\section{Introduction}

 Relativistic jets in  radio-loud active galactic nuclei
 (AGNs)
 dissipate their kinetic energy via
 interactions with surrounding  interstellar medium (ISM) or
 intracluster medium (ICM), 
 and inflate a bubble composed of decelerated jet matter,
 which is often referred to as cocoon. %
 Initially, the cocoon is highly overpressured against 
 the ambient ISM/ICM  and a strong  shock is driven 
into the ambient matter. Then a thin shell is formed around the cocoon 
by the compressed ambient medium.
 As in other astrophysical shocks, %
 the shells are expected to be a promising site for particle accelerations, 
  since the shocks are driven into tenuous plasmas.
 In the present study, we  explore  the 
evolution of the non-thermal emissions by the accelerated electrons in 
the shocked shells.
 We
 properly take into account the Comptonization of photons
 of various origins
 which were not considered in the previous studies.\cite{FKY07} 
 Focusing on the powerful sources
 which host luminous quasar in its core,
 we show, in particular, that the  energy
 of accelerated electrons is efficiently converted through the IC scattering to
 high energy $\gamma$-rays of up to $\sim 10~{\rm TeV}$ if the source is 
relatively compact.

\section{Model}

%
  When considering the dynamics
 of the expanding cocoon and shell, 
 we neglect the elongation 
 in the jet direction 
  and assume that they are spherical for simplicity.
 We also assume that 
 the ambient mass density profile has a form of a power-law
 given by
 $\rho_a(r) = \rho_{0} (r / {\rm 1kpc})^{-1.5}$.
 We further assume that the kinetic power of jet, $L_{\rm j}$,
 is constant in time. 
%
 Under these assumptions,
 the dynamics can be approximately described based on the model 
 of  stellar wind bubbles.\cite{CMW75}
 Then radius of the shock is written as 
 $R(t) \sim  22  {\rho}_{0.1}^{-2/7} L_{45}^{2/7}
                   t_{7}^{6/7}~
                   {\rm kpc} $,      
  where
  $\rho_{0.1} = \rho_0 / 0.1 m_p ~{\rm cm}^{-3}$,
  $L_{45} = L_{\rm j}/10^{45}~{\rm ergs~s^{-1}}$ and
  $t_{7} = t / 10^7~{\rm yr}$. 
  Also the total internal energy stored in the shell
  can be expressed as
  $E_{\rm s} \sim 0.1 L_{\rm j} t$,
 implying that roughly $10\%$ of the total energy released by the 
  jet is deposited in the shell.

 The energy 
 distribution of the non-thermal electrons is determined 
 by solving the kinetic equation in one-zone approximation
 which takes in account the injection of electrons and
 the cooling effects.\cite{MK07}
 The electron injection rate $Q(\gamma_e)$ and the cooling rate
 $\dot{\gamma}_{\rm cool}$,
 which will be described  below,
 are evaluated based on the dynamical model described above.

%
 We assume that the  electrons are injected into
 the post-shock region with a power-law energy distribution
 given as
$Q(\gamma_e) =
                K \gamma_e^{-2}$
         (for $1\leq \gamma_e
                       \leq \gamma_{\rm max}$),
 where
  $\gamma_{\rm max}$ corresponds to
 the maximum Lorentz factor.
 The value of $\gamma_{\rm max}$ is obtained by equating the 
 the cooling rate, $\dot{\gamma}_{\rm cool}$,
 to the  acceleration rate given by 
  $\dot{\gamma}_{\rm accel} = (3/20)( e B \dot{R}^2 / \xi  m_e c^3)$, 
 where $B$
 and $\dot{R}$ are
 the magnetic field strength in the post-shock region and
 the expansion velocity of the shell, respectively.
 Here, $\xi$ is the so-called ``gyro-factor'' which can be identified with  the ratio of the energy
 in ordered magnetic fields to that in turbulent ones.
 We postulate $\xi \sim 1$ (Bohm limit), as is observed to be
 the case for some SNRs. \cite{YYT04}\cdash\cite{SAH06}
 Assuming that the magnetic field of ambient ISM/ICM
 ($\sim$ few ${\rm \mu G}$)\cite{MS96}\cdash\cite{SCK05}
 is adiabatically compressed by the shock,
 we take $B = 10{\mu {\rm G}}$ as a fiducial value
 for the magnetic field strength.
 The normalisation factor, $K$,  is determined from the assumption 
 that a fraction, $\epsilon_e$, of the  energy stored in the shell
 is carried by the 
 non-thermal electrons.
 In the present study, as a fiducial case, we assume  $\epsilon_e = 0.01$.
 It is noted that since 
 the factor $K$ is proportionate to $\epsilon_e$ and $L_{\rm j}$, 
 the resultant luminosity of non-thermal emissions also scales in the same 
 manner with these quantities.

 In the cooling rate, $\dot{\gamma}_{\rm cool}$,
 the adiabatic losses due to  expansion of the shell and
 the radiative losses due to synchrotron and IC emissions
 are taken into account.
 Regarding the synchrotron losses, the magnetic field considered above 
 is used.
 In evaluating  the cooling rate for IC scattering, 
 we take into account various seed photons of relevance in this context. 
 The considered photon fields are
 UV emissions 
 from the accretion disc, IR emissions from the dusty torus,
 stellar emissions from the host galaxy in NIR,
 synchrotron emissions from the radio
 lobe and CMB.  
 We assume that the photons from the disc,  torus,  host galaxy,
 and CMB are monochromatic and have the following single frequencies:
 $\nu_{\rm UV} = 2.4 \times 10^{15}~{\rm Hz}$, 
 $\nu_{\rm IR} = 1.0 \times 10^{13}~{\rm Hz}$, 
 $\nu_{\rm NIR} = 1.0 \times 10^{14}~{\rm Hz}$, 
 and
$\nu_{\rm CMB} = 1.6 \times 10^{11}~{\rm Hz}$.
The photons from the radio lobe are assumed to have a continuous spectrum 
given by $L_{\rm \nu, lobe} \propto \nu^{-0.75}$.
 In the present study, we focus on powerful sources
 hosting  luminous quasars and
 adopt $L_{\rm UV} = 10^{46}~{\rm ergs~s^{-1}}$ 
 for the luminosity  of the UV emissions from the disc.
 We assume  that 
 the luminosity of the IR emissions from the torus
 is equal to that of the UV emissions
 ($L_{\rm IR}=L_{\rm UV}$). 
 The luminosity of the host galaxy is assumed as
 $L_{\rm NIR} = 10^{45}~{\rm ergs~s^{-1}}$.
 Finally, the luminosity of the lobe is
 determined by assuming that
 a fraction $\eta$ of the
 jet power is radiated as radio emissions from the lobe
 (i.e., $L_{\rm  lobe} = \eta L_{\rm j}$).
 Here we assume $\eta = 10^{-2}$ as a fiducial case.

\section{Non-thermal Emissions}

\begin{figure}[pb]
\centerline{\psfig{file=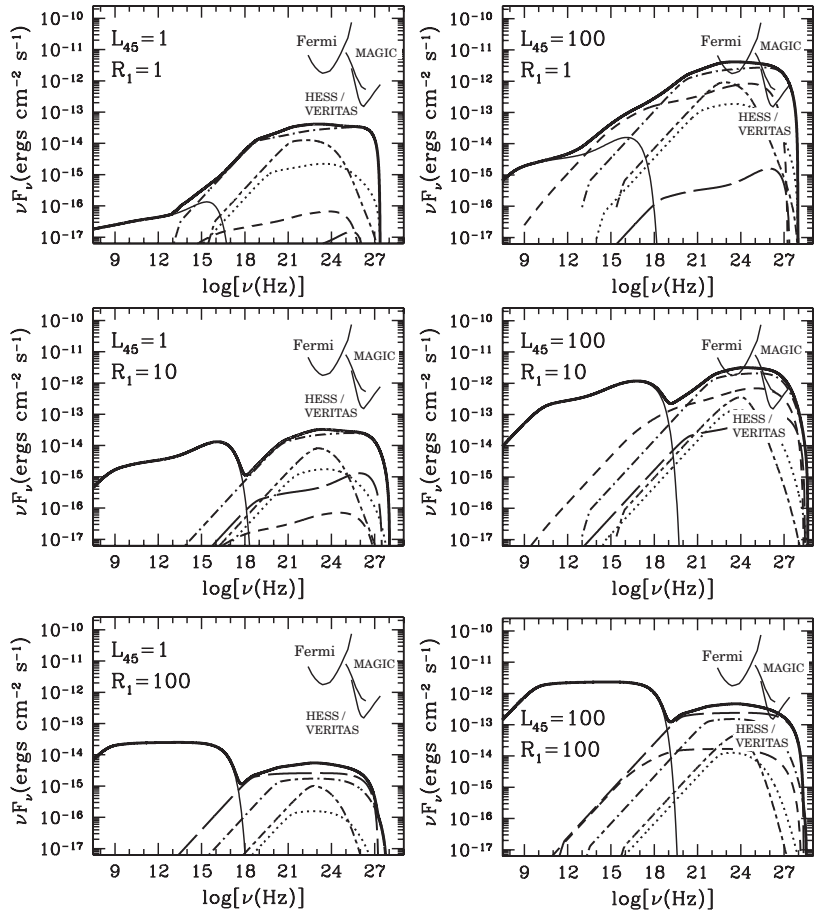,width=12cm}}
\vspace*{8pt}
\caption{The spectra of the synchrotron and IC emissions
 from sources with the jet powers of
 $L_{\rm j} = 10^{45}{\rm ergs~s^{-1}}$ ({\it left panels})
 and $L_{\rm j} = 10^{47}{\rm ergs~s^{-1}}$ ({\it right panels})
 located at the distance of $D=100~{\rm Mpc}$.
 The top, middle and bottom panels are displayed for the source 
 sizes of $R =1$, $10$, and $100~{\rm kpc}$, respectively.
 The various lines show the contributions from the synchrotron 
 emissions ({\it thin solid line}) and IC scatterings of
 UV disc photons ({\it long-short-dashed line}),
 IR torus photons ({\it dot-dashed line}),
 NIR host-galaxy photons ({\it dotted line}),
 CMB photons ({\it long-dashed line}) 
 and lobe photons ({\it short-dashed line}).
 The thick solid line is the sum of these fluxes.
%
 \label{f1}}
\end{figure}

 From the obtained electron distribution, we calculate the 
 spectra of synchrotron and IC radiations.
  In Fig. \ref{f1}  we show the photon fluxes,
 $\nu F_{\nu}$, for
 sources located at distance of  $D = 100~{\rm Mpc}$.
The left panels show the case for sources with
 jet powers of $L_{\rm j} = 10^{45}~{\rm ergs~s^{-1}}$,
while the right panels show the case for
   $L_{\rm j} = 10^{47}~{\rm ergs~s^{-1}}$.   
 The top, middle and bottom panels of the figure correspond to the source sizes of
 $R = 1~{\rm kpc}$, $10~{\rm kpc}$ and $100~{\rm kpc}$, respectively.
 In addition to the total photon flux ({\it thick solid line}),
 we show
 the contributions from the synchrotron emissions ({\it thin solid line})
 and the IC scatterings of UV disc photons ({\it long-short-dashed line}),
 IR torus photons ({\it dot-dashed line}), NIR host-galaxy photons ({\it dotted line}),
 CMB photons ({\it  long-dashed line}) and lobe photons ({\it short-dashed line}).
%

 The synchrotron emissions are the main low-frequency component,  
 which extends from radio to $\sim {\rm keV}$ X-ray.
 The IC emissions become remarkable at higher frequencies 
 up to $\sim 10 ~{\rm TeV}$ gamma-ray.
 When the source is young and hence small, the radiative
 output is dominated by the IC emissions, since the energy density of 
 photons  is larger than that of magnetic fields
 (referred to as the IC-dominated stage).
 As the the source becomes larger, on the other hand, 
 the energy density of photons decreases ($U_{\rm ph} \propto R^{-2}$)
 and the synchrotron emissions becomes dominant
 (the synchrotron-dominated stage).  
Among the contributions to the IC emissions,
 the scattering of the IR torus photons 
 is the largest at least in the IC-dominated stage
 thanks to the high energy density of the
 IR photons. 
 Note that, 
although the UV disc photons are assumed to have the same energy density as 
the IR photons,
 the IC scattering of UV photons 
 is suppressed at the frequencies above
 $\nu \gtrsim 10^{24}~{\rm Hz}$ by the Klein-Nishina effect.
 Then the transition 
 from the IC-dominated stage to the synchrotron-dominated stage
 occurs roughly at
 $R_{\rm IC/syn} \sim 27 L_{\rm IR, 46}^{1/2} B_{-5}^{-2}{\rm kpc}$
 which corresponds to the condition $U_{\rm IR} \sim U_{\rm B}$.
%
 While the contributions from the UV disc photons,
  host-galaxy photons and
 lobe photons 
 are modest at best through the entire evolution, 
 the IC scattering of CMB photons 
 dominates over other IC components
 for sources larger than  $R \sim 85 L_{\rm IR, 46}^{1/2}{\rm kpc}$.

 The peak luminosities in the spectra
 are roughly equal to the energy injection rate
 on the non-thermal electrons
 ($\gamma_e^2 m_e c^2 Q(\gamma_e) \sim \nu L_{\nu}$) 
 because the cooling time scale of the high energy electrons
 is shorter than the dynamical time scale (fast cooling). 
 Since the energy injection rate is independent of the electron energy 
 ($\gamma_e^2 Q(\gamma_e) \propto \gamma_e^0$),
  these non-thermal electrons produce a rather flat and broad spectrum
 ($\nu L_{\nu} \propto \nu^{0}$) in the corresponding frequency range.
 We can give a rough estimate to
 the peak luminosity as
 $\nu L_{\rm \nu, peak}
  \sim  5.0 \times 10^{40} \epsilon_{-2}L_{45} 
~{\rm ergs~s^{-1}}$,
 or, equivalently, to the peak flux as 
 $\nu F_{\rm \nu, peak} \sim  4 \times 10^{-14}
   \epsilon_{-2}L_{45}D_{2}^{-2}
~{\rm ergs~cm^{-2}s^{-1}}$,
 where $\epsilon_{-2} = \epsilon_e / 0.01$ and $D_2 = D/100~{\rm Mpc}$.
%
 The feature is clearly seen in Fig. \ref{f1}. 
Indeed, the spectra are flat with
 a peak flux given approximately by the above estimation.
 As mentioned in the previous section,
 the emission luminosity 
 scale approximately linearly
 with the acceleration efficiency $\epsilon_e$ and  the
 jet power $L_{\rm j}$.
 For given values of $\epsilon_e$ and $L_{\rm j}$,
 while the value of $\nu L_{\rm \nu, peak}$ remains nearly constant,
the frequency range, where the spectrum is flat, varies with the source size
because of the changes in the energy range of the fast cooling electrons
and the main emission mechanism (synchrotron or IC). 
 It is emphasized that the peak luminosity
 is chiefly governed by $\epsilon_e$ and $L_{\rm j}$
 and is quite insensitive to
 the  magnetic field strength and seed photons, which will
 only affect the frequency range of the flat spectrum.
%
 This means that if $L_{\rm j}$  is constrained by other independent
 methods,
\cite{AFT06}\cdash\cite{IKK08}
 the observation of the peak luminosity will enable us to obtain information on the 
acceleration efficiency $\epsilon_e$.

 Next we consider the detection prospect. The synchrotron emissions can be 
 observed at frequencies from radio to X-ray. Obviously large sources ($R \gtrsim R_{\rm IC/syn}$) 
 offer a greater chance of detection than small ones,
 since the synchrotron radiation is strongly suppressed in the compact sources
 and the small spatial scale will make it difficult to distinguish the synchrotron radiations 
 from the core emissions of AGN.
 Even for large sources ($R \gtrsim R_{\rm IC/syn}$), however, 
 the synchrotron emissions are subject to contaminations with 
 radio emissions from the lobe, optical emissions from the host galaxy
 and X-ray emissions from ISM/ICM, which are at least partially cospacial.
 The obtained luminosity of the shell emissions is likely to be 
lower than that of these emissions.
 Hence the observation of the synchrotron emissions will be very difficult irrespective of the
source size.

 In the case of the IC emissions, which are pronounced in gamma-ray,
 compact sources ($R \lesssim R_{\rm IC/syn}$) are favored for detection, 
 since the luminosity is higher. 
 Also significant contamination is not expected in this energy range
for non-blazar AGNs. 
%
%
 At the photon energy of $h \nu \sim {\rm GeV}$,
 the detection limit of the Fermi gamma-ray telescope
 is roughly $\sim 10^{-12}~{\rm ergs~cm^{-2}~s^{-1}}$ whereas
 modern Cherenkov telescopes such
 as HESS, MAGIC and VERITAS have a detection limit of 
 $\sim 10^{-13}~{\rm ergs~cm^{-2}~s^{-1}}$ at $h \nu \sim {\rm TeV}$.
 From the estimated peak flux, $\nu F_{\rm \nu, peak}$,
 we find that the currently operating gamma-ray telescopes 
 are capable of detecting these emissions 
 at $\sim {\rm GeV}$ and $\sim {\rm TeV}$ 
 if the jet power satisfies $L_{\rm j} \gtrsim 3 \times
 10^{46}\epsilon_{-2}^{-1}D_{2}^{2} ~
  {\rm ergs~s^{-1}}$ 
 and
 $L_{\rm j} \gtrsim 3 \times 10^{45}\epsilon_{-2}^{-1}D_{2}^{2}
 ~ {\rm ergs~s^{-1}}$, respectively, and the source size is smaller than $R_{\rm IC/syn}$.
 This can be confirmed in 
 Fig. \ref{f1} indeed.
 For the most powerful source with the jet power of $L_{\rm j} = 10^{47}~{\rm ergs~s^{-1}}$
located at  $D = 100~{\rm Mpc}$,
 $\sim {\rm GeV}-{\rm TeV}$ gamma-rays from the shell may be accessible to the 
 Fermi, MAGIC, HESS and VERITAS gamma-ray telescopes 
 if the source is compact. 
%

\section{Summary}
\label{summ}

 We have explored the temporal evolution of the emissions by accelerated electrons in 
 the shocked shell produced by AGN jets. 
 Below we summarize our main findings in this study.

 \noindent (i) 
 When the source is young and small ($R \lesssim R_{\rm IC/syn} \sim  27 L_{\rm IR, 46}^{1/2} B_{-5}^{-2}{\rm kpc}$),
 the dominant radiative process is the IC scattering of IR photons emitted 
 from the dust torus.
 For larger sources, on the other hand, the synchrotron emissions dominate 
over the IC emissions, since the energy density of photons becomes smaller than 
that of magnetic fields ($U_{B} > U_{\rm ph} \propto R^{-2}$).
Through the entire evolution, the spectrum is rather broad and flat, and 
the peak luminosity is approximately given by 
 $\nu L_{\rm \nu, peak} \sim  3.0 \times 10^{40} \epsilon_{-2}L_{45}~{\rm ergs~s^{-1}}$, 
since it is roughly equal to the energy injection rate, which is in turn determined 
by the jet power $L_{\rm j}$ and acceleration efficiency $\epsilon_e$. 

\noindent (ii)
 The spectra of the
  IC emissions extend up to 
  $\sim 10~{\rm TeV}$ gamma-ray energies
 for a wide range of source size ($R\sim 1-100~{\rm kpc}$)
 and jet power ($L_{\rm j}\sim 10^{45}-10^{47}~{\rm ergs~s^{-1}}$).
 For most powerful nearby sources  ($L_{\rm j} \sim 10^{47}~{\rm ergs~s^{-1}}$, 
$D \lesssim 100~{\rm Mpc}$), ${\rm GeV}-{\rm TeV}$  gamma-rays produced via
the IC emissions can be detected by Fermi/LAT as well as by the modern Cherenkov telescopes
 such as MAGIC, HESS and VERITAS if the source is compact ($R \lesssim R_{\rm IC/syn}$). 
  The observation of these emissions  enable us to
 probe the acceleration efficiency  $\epsilon_e$ of which
 little has been known so far.

\section*{Acknowledgments}

 This study was partially supported by the Grants-in-Aid for the
 Scientific Research (17540267, 19104006, 21540281) from Ministry of Education,
 Science and Culture of Japan and 
 by Grants-in-Aid for the 21th century
 COE program ``Holistic Research and Education Center for Physics of
 Self-organizing Systems''.
 This work was supported by
 Research Center for the Early Universe.
 NK is financially supported by the Japan Society for the Promotion
 of Science (JSPS) through the JSPS Research Fellowship for
 Young Scientists.
%


\end{document}